\documentclass[aps,prl,twocolumn,showpacs,groupedaddress,nofootinbib]{revtex4}
\usepackage{graphicx}
\usepackage{hyperref}
\usepackage{color}

\begin{document}
\title{Distribution Principle of Bone Tissue}

\author{Yifang Fan$^{a}$\footnote{Corresponding author\\
Email: tfyf@mailer.gipe.edu.cn}, Mushtaq Loan$^{b}$,
 Yubo Fan$^{c}$, Zongxiang Xu$^{a}$ and Zhiyu Li$^{d}$}
\affiliation{$^{a}$Center for Scientific Research, Guangzhou
Institute of Physical Education, Guangzhou 510500, China\\
$^{b}$International School, Jinan University,Guangzhou 510632, China\\
$^{c}$Bioengineering Department, Beijing University of
Aeronautics and Astronautics, Beijing 100191, China\\
$^{d}$College of Foreign Languages, Jinan University, Guangzhou
510632, China}

\date{\today}

\begin{abstract}
Using the analytic and experimental techniques we present an
exploratory study of the mass distribution features of the high
coincidence of centre of mass of heterogeneous bone tissue in vivo
and its centroid of geometry position. A geometric concept of the
average distribution radius of bone issue is proposed and functional
relation of this geometric distribution feature between the
partition density and its relative tissue average distribution
radius is observed. Based upon the mass distribution feature, our
results suggest a relative distance assessment index between the
center of mass of cortical bone and the bone center of mass and
establish a bone strength equation. Analysing the data of human foot
in vivo, we notice that the  mass and geometric distribution laws
have expanded the connotation of Wolff's law, which implies a leap
towards the quantitative description of bone strength. We finally
conclude that this will not only make a positive contribution to
help assess osteoporosis, but will also provide guidance to exercise
prescription to the osteoporosis patients.
\end{abstract}

\pacs{87.85.G-, 87.85.J-, 06.30.Dr.\\
 Keywords: Wolff's law, max-min principle, osteoporosis, bone strength}

\maketitle

Bone tissue structure and function are largely associated with its
mechanical and biological environment
\cite{Wolff,Chen,Rubin2001,Marco,Norbert}. Growth, modeling and
remodeling are the basic physiological features of bone.
Biomechanically, bone growth is defined as mass changes \cite{Fung},
and mechanical force and movement play a role in bone growth
\cite{Buckwalter}. When the skeleton bears loads externally, bone
tissue will undergo adaptive changes such as reabsorption or
remodeling \cite{Burger}, and point-to-point changes to the material
property by changing mass distribution
\cite{Harrigan1984,Odgaard,Bagge} will maximize its external loads.
Individual bone growth indicates that external force has great
effect on cross-sectional geometry and internal anatomy \cite{Ruff}.
Bone structure is an optimization of stress transformation
\cite{Wolff} and it is an adaptive response to incorporation of
minimal weight to maximal strength by some special rules
\cite{Roesler}. Bone physiological activity is regarded as a process
of optimization \cite{Turner1992, Huiskes, Harrigan, Marco}.
Consequently, stress has caused adaptive changes of bone shape and
structure, which involve constant optimization of structures. But it
remains unclear what distribution principle these changes follow.

From the biomechanical perspective, osteoporosis means a sharp drop
of bone mass and strength and they cannot meet the demands of
adaptive strength and movement load \cite{Frost}. Many mechanical
models adequately represent the relation between the bone geometry
and its strength \cite{Gemunu, Weinkamer, Rubin2005}, as well as the
correlation between bone density and its strength \cite{Crawford}.
While the phenomenological models may often be helpful in obtaining
a qualitative understanding of the data, microscopically these
models do not provide a trustworthy guide into unknown territory of
an accurate relation between the distribution of bone tissue and its
strength. Our approach will be to use a combination analysis of
analytic and experimental techniques to examine the the above two
uncertain areas by setting up a bone strength equation to obtain a
quantitative description macroscopically.

Centroid of geometry (hereinafter referred to as COG) and the center
of mass (hereinafter referred to as COM) of the homogeneous
materials are coincide, whereas in most cases those of heterogeneous
do not. Using CT (computed tomography) scan technology, we conducted
the analysis to explore the relation between COM and COG of bone in
the physiological activities, such as continuous modeling and
remodeling in its adaptive mechanical condition. At small enough CT
resolution rate and its slice distance, the bone tissue density of
infinitesimal bone volume segmentation, $dV$, can be regraded as
continuous. Its point density $\rho_{i}$ can be taken as that of the
micro-element, $dm_{i}=\rho_{i}dV$, thereby approximating the bone
as a collection of  particles. To find the optimal program for
heterogeneous bone tissue mass distribution, we minimize
\begin{equation}
\min \Psi(p_{c})=\sum\rho_{i}\Delta
V\bigg((x_{i}-x_{c})^{2}+(y_{i}-y_{c})^{2}+(z_{i}-z_{c})^{2}\bigg),
\label{eqn1}
\end{equation}
where $p_{c}(x_{c},y_{c},z_{c})$ and $p_{i}(x_{i},y_{i},z_{i})$
refer to the relative locations of coordinators of bone COM and
random point related to CT image, respectively. The series
$\sum\rho_{i}\Delta V, \sum\big(\mid x_{i}-x_{c}\mid +\mid
y_{i}-y_{c}\mid + \mid z_{i}-z_{c}\mid\big), \sum\rho_{i}\Delta
V\big(\mid x_{i}-x_{c}\mid +\mid y_{i}-y_{c}\mid + \mid
z_{i}-z_{c}\mid\big)$ are all convergent. Thus in the limit $\Delta
V\rightarrow 0$ $(\Delta V=abc$, $a\rightarrow 0$, $b\rightarrow 0$
and $ c\rightarrow 0 )$, the Abelian theorem on series \cite{Hardy}
leads to
\begin{equation}
\overline{p}=p_{c}, \label{eqn2}
\end{equation}
where we have used $\overline{p}(\overline{x},\overline{y},
\overline{z})$ for the COG. This shows that the optimal program for
the heterogeneous bone tissue mass distribution should be the
coincidence of its COM and COG. This signature does not, however,
indicate a similar behaviour of the tissue distribution of bone in
vive. In order to study the relation between the COM and COG of bone
in vive, a CT scanning\footnote{The scanning is done by
Philips/Brilliance 64 ($120kv$, Pixel Size: $0.328-0.475mm$, Slice
Distance $0.330-0.450mm$).} is conducted to the foot of eight
volleyballers, eight classical wrestlers and two senior females.
\begin{table}[tbp]
\scriptsize \caption{ \label{tab1} Basic Information of the
subjects.}
\begin{ruledtabular}
\begin{center}
\begin{tabular}{cccccccc}
       & Wrestlers & Volleyballers  & Seniors\\ \hline
Sample size & $8$ & $8$ & $2$\\
Age($year$) & $21.00\pm 2.78$ & $21.88\pm 0.99$ & $64.50\pm 4.95$ \\
Height($cm$) & $168.00\pm 5.68$ & $183.94\pm 3.90$ & $150.50\pm 3.54$ \\
Body mass($kg$) & $65.52\pm 5.16$ & $71.80\pm 5.20$ & $52.88\pm 3.15$\\
Calcaneus volume($cm^{3}$)  & $71.79\pm 7.86$ & $81.79\pm 4.26$ & $49.43\pm 5.22$ \\
Calcaneus density($g/ml$) & $1.47 \pm 0.04 $ & $1.49 \pm 0.05 $ & $1.28 \pm 0.03 $ \\
\end{tabular}
\end{center}
\end{ruledtabular}
\end{table}
\normalsize The position of COG is calculated by
\begin{displaymath}
\frac{\sum x}{\sum 1}, \frac{\sum y}{\sum 1},\frac{\sum z}{\sum 1}
\end{displaymath}
and that of COM by
\begin{displaymath}
\frac{\sum x\rho}{\sum \rho},\frac{\sum y\rho}{\sum \rho},\frac{\sum
z\rho}{\sum \rho}
\end{displaymath}
The comparative testing accurate distance between COM and COG
positions is evaluated using
\begin{displaymath}
\frac{\sqrt{(\overline{x}-x_{c})^{2}+(\overline{y}-y_{c})^{2}+(\overline{z}-z_{c})^{2}}}{\sqrt{a^{2}+b^{2}+c^{2}}}
\end{displaymath}
\begin{figure}[tbp]
\scalebox{1}{\includegraphics{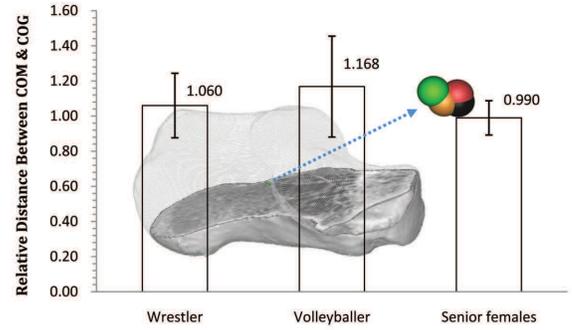}} \caption{ \label{fig1} The
position relation between COM and COG. The green ball refers to the
COG of calcaneus while the red, black and orange balls represent
positions of calcaneus COM from the wrestlers, volleyballers and the
senior females (the colored ball radii are all
$\frac{1}{2}\sqrt{a^{2}+b^{2}+c^{2}}$).}
\end{figure}
where $a=\frac{1}{X}$, $b=\frac{1}{Y}$ $XY$ the CT image resolution,
$c$ the slice distance.

Fig. \ref{fig1} collects and displays our data shown in Table
\ref{tab1}. The behaviour seen in Fig. \ref{fig1} confirms our
signature that the high coincidence of the position of COM and COG
of heterogeneous bone in vive is independent of the changing process
of bone in its adaptive mechanical environment and that the bone
tissue mass distribution observes the optimal principle with a
coincidence between COM and COG.

The above signature brings us to the issue of mass distribution
index of bone tissue. The mechanical properties reveal that the
elastic property and pressure-bearing strength of cortical bone is
several times more than those of the same-volume spongial bone
\cite{Buckwalter}. Using the mechanical insight, we divide the
tissue continuous density into three parts; bone marrow, spongial
bone and compact bone with their COM and COG highly coincident as
indicated by the observed fact ${\bar p} =p_{c}$. Using
$\sqrt{(x_{ci}-x_{c})^{2}+ (y_{ci}-y_{c})^{2}+(z_{ci}-z_{c})^{2}}$,
where $(x_{ci},y_{ci},z_{ci})$ refer to COM of bone tissue and
$(x_{c},y_{c},z_{c})$  refers to COM of calcaneus, we  calculate the
distance between each individual tissue COM and that of the
calcaneus. In order to avoid the effect from the size of the
subject's calcaneus, we standardize the average distribution radius
of calcaneus so as to compare calcaneus of different volume, thereby
establishing an anastz for the distribution index:
\begin{equation}
ID=\frac{\sum1\sqrt{(x_{ci}-x_{c})^{2}+(y_{ci}-y_{c})^{2}
+(z_{ci}-z_{c})^{2}}}{\sum(\sqrt{(x_{j}-x_{c})^{2}+(y_{j}-y_{c})^{2}+(z_{j}-z_{c})^{2}})},
\label{eqn3}
\end{equation}
where $i=1,2,3$ stands for bone marrow, spongial bone and compact
bone respectively. $j$ refers to all the tissues that make up bone
and $c$ stands for COM.

\begin{figure}[tbp]
\scalebox{1}{\includegraphics{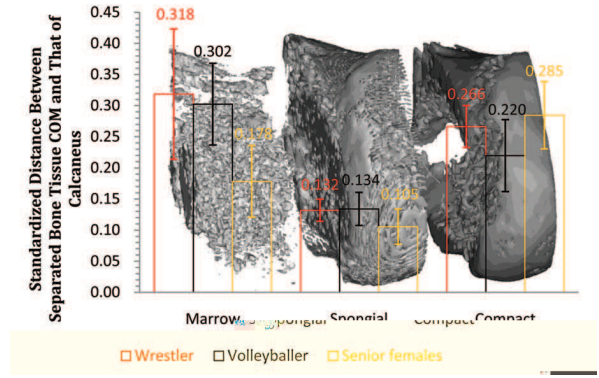}} \caption{ \label{fig2}
Relative positions of bone tissue COM and the calcaneus COM.}
\end{figure}

From Fig. \ref{fig2} it is clear that the position of compact bone
COM of the volleyballers is closest to that of the calcaneus COM.
Distinct difference exists between the volleyballers and wrestlers,
so is the difference between the senior females and the
volleyballers and wrestlers. The distance of the senior females is
the largest. If such a trend continues for larger sample sizes, it
will add one more quantitative evaluation index while diagnosing
osteoporosis.

To see whether the changes of BMC (bone mineral content) and BMD
(bone mineral density) bring about changes in the geometric
distribution of bone tissue simultaneously, we segment the density
$\rho_{i}$ and use $\overline{r}_{i}=\frac{\sum \Delta r}{\sum1}$ to
calculate the average distribution radius of the relative tissue of
segmented density to the calcaneus COM. When the continuity of
segment density is guaranteed and the calcaneous density variation
ranges are fixed, the relation between the calcaneous density and
the average bone tissue distribution radius has been developed.
After an analysis of the fitting function, we establish a functional
relationship between the two for a typical value of correlation
coefficient $R>0.99$.
\begin{figure}[tbp]
\scalebox{1}{\includegraphics{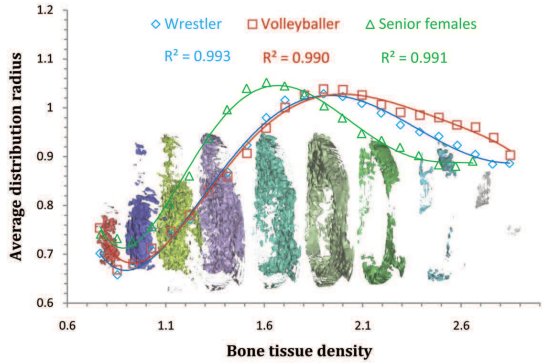}} \caption{ \label{fig3}
Relation between bone tissue density and its average distribution
radius. Density is converted by
$\rho=\frac{\rho_{GrayValues}}{\rho_{H_{2}O}}$ and distribution
radius standardized by $\rho_{i}$.}
\end{figure}
Fig. \ref{fig3} shows the obvious  difference of average
distribution radii between the senior females and the athletes. For
the density $>1.7$, the average distribution radius of the senior
females drops dramatically. When comparing the wrestlers and the
volleyballers, when the density is greater than 2.0, the bone tissue
average distribution radius of the volleyballers is more than that
of the wrestlers. That is to say, when the geometrical distribution
of bone tissue follows Eq. (\ref{eqn2}), senior females' bone loss
will be accompanied by a decrease of distribution radius of compact
bone.  When comparing the volleyballers and wrestlers, there will be
an apparent increase of distribution radius of compact bone. We may
safely say that the bone adaptability includes not only changes of
BMC and BMD, but also  changes in radius of bone tissue
distribution. Once the bone shape, tissue density and their
corresponding volume have been determined, the tissue geometric
distribution will determine the strength of bone tissue. The moment
of inertia of bone is an important index to reflect bone strength.
For the heterogeneous materials the bone density and intensity
follow a non-linear relationship. We introduce a coefficient
$e^{\rho^{k}}$ and combine the segmented density and intensity to
calculate the segmented strength on the basis of moment of inertia

\begin{displaymath}
M_{i}r^{2}_{i}e^{\rho_{i}^{k}}=\rho_{i}V_{i}r^{2}_{i}e^{\rho_{i}^{k}}
\end{displaymath}
and establish a functional relation between the calcaneus bone
strength and the segmented density as
\begin{equation}
\sigma=\int^{b}_{a}f(\rho)d\rho, \label{eqn4}
\end{equation}
where $\sigma$ refers to strength of calcaneus, $f(\rho)=\rho
Vr^{2}(\rho)e^{\rho^{k}}$. When $\Delta\rho$ is small enough, and
$f(a)\neq f(b)$, Eq. \ref{eqn4} can be approximated by
\begin{displaymath}
\sigma=(\rho_{max}-\rho_{min})\frac{\sum
\rho_{i}V_{i}r^{2}_{i}\exp(\rho^{k}_{i})}{\sum1}.
\end{displaymath}
$\rho_{max}$ and $\rho_{min}$ refer to the maximal value of compact
bone density and the minimal value of spongial bone density,
respectively and $\exp(\rho^{k}_{i})$ is the coefficient parameter
of bone density and strength. Note that the above equation holds for
continuous heterogeneous material only.

\begin{figure}[tbp]
\scalebox{1}{\includegraphics{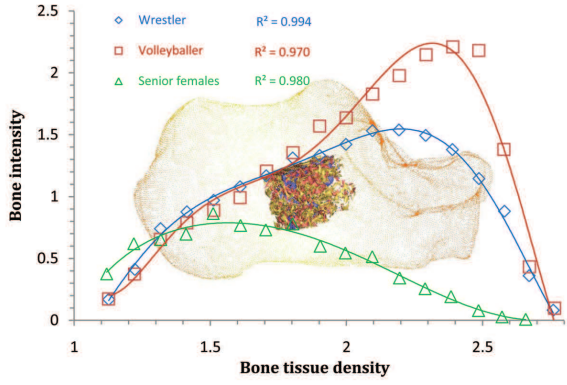}} \caption{ \label{fig4}
Geometric distribution of bone strength.  The tissue volume of the
segmented density has been normalized by $V_{i}=\frac{\sum \Delta
V}{V}$, and its distribution radius standardized by
$\overline{r}_{i}=\frac{\sum \Delta r}{\overline{r}\sum1}$. The
typical value of parameter $k$ is chosen to be $1.68$.}
\end{figure}

Fig. \ref{fig4} shows that when the bone tissue density of wrestlers
and volleyballers is greater than 1.8, the difference in intensity
between the two grows and reaches its maximum in the range of $2.4 -
2.5$. On the other hand, the bone intensity  of the senior females
begins to show a larger difference with that of the athletes for
density $> 1.4$.

Concerning bone tissue distribution, our study confirms that the
high coincidence of COM and COG of heterogeneous material. This
coincidence is the prerequisite to meet the requirement of
max-min-principle and forms the bases for developing an evaluation
index. We noticed that there is no difference in spongial bone
between the wrestlers and volleyballers, whereas there is obvious
difference in compact bone. This would explain the movement of the
compact bone COM towards the calcaneus COM, which enables the
calcaneus structure to bear greater stress. This can also be a
representation that bone can yield adaptive changes functionally.
What's more significant is the fact that the compact bone COM of the
senior females moves away from the calcaneus COM. If a larger size
sample can verify this, it will bring greater significance to the
clinical practice.

One of the main aims of bone study is to conduct qualitative
analysis to bone strength. Bone strength relies on bone mass, but
the uni-index of bone mass cannot paint a holistic and realistic
picture of bone intensity objectively \cite{Crawford, Rubin2005,
Bagi}. Modeling analysis reveals that the main factors that
determine the structure strength include mass distribution,
geometric distribution and moment of inertia of various tissues
\cite{Hudelmaier, Mittra}. Eq. (\ref{eqn4}) has successfully
combined those factors and from a mathematical perspective it has
illustrated that volume, mass, density, distribution radius and
moment of inertia of bone tissue cannot be employed individually to
assess bone strength.  Eq. (\ref{eqn4}) also reveals that criteria
of selecting a training approach to increase or improve bone mass of
compact bone and its distribution radius that are clinically
significant. The bone tissue geometric distribution principle sheds
light on the effect to bone structure from different types of
training. A geometrically-distributed bone strength equation can
mirror the effects to bone strength from various factors. When the
physiological bone mass decrease has become unavoidable, will
exercise patterns be able to change its geometric distribution? If
yes, Eq. (\ref{eqn4}) will undoubtedly provide some guidance to the
development of exercise prescription and it can also be employed as
an important evidence to examine and modify exercise prescription.
\section{ acknowledgment}
This project was funded by  National Natural Science Foundation of
China under the grant $10772053$ and by Key Project of Natural
Science Research of Guangdong Higher Education Grant No $06Z019$ .
The authors would like to acknowledge the support from the subjects.

\end{document}